\documentclass[12pt]{spieman}  
\usepackage{amsmath,amsfonts,amssymb}
\usepackage{graphicx}
\usepackage{setspace}
\usepackage{tocloft}
\usepackage{lineno}
\usepackage{color,soul}
\usepackage[dvipsnames]{xcolor}
\usepackage{multicol}
\usepackage{siunitx}
\usepackage[all]{nowidow}
\usepackage{booktabs}
\DeclareRobustCommand{\rchi}{{\mathpalette\irchi\relax}}
\newcommand{\irchi}[2]{\raisebox{\depth}{$#1\chi$}} 

\title{Comparative laboratory study of electric field conjugation algorithms}

\author[a*]{Niyati Desai}
\author[b,c]{Axel Potier}
\author[a]{Susan F. Redmond}
\author[b]{Garreth Ruane}
\author[b]{Phillip K. Poon}
\author[b]{A J Eldorado Riggs}
\author[b]{Matthew Noyes}
\author[b] {Camilo Mejia Prada}
\affil[a]{Department of Astronomy, California Institute of Technology, 1200 East California Blvd., Pasadena, CA, 91125}
\affil[b]{Jet Propulsion Laboratory, California Institute of Technology, Pasadena, CA 91109, USA}
\affil[c]{Division of Space and Planetary Sciences, University of Bern, Sidlerstrasse 5, 3012 Bern, Switzerland}

\cftpagenumbersoff{figure}
\cftpagenumbersoff{table} 
\begin{document} 
\maketitle

\begin{abstract}

Future space telescope coronagraph instruments hinge on the integration of high-performance masks and precise wavefront sensing and control techniques to create dark holes essential for exoplanet detection. Recent advancements in wavefront control algorithms might exhibit differing performance depending on the coronagraph used. This research investigates three model-free and model-based algorithms in conjunction with either a vector vortex coronagraph or a scalar vortex coronagraph under identical laboratory conditions: pairwise probing with electric field conjugation, the self-coherent camera with electric field conjugation, and implicit electric field conjugation. We present experimental results in narrowband and broadband light from the In-Air Coronagraph Testbed at the Jet Propulsion Laboratory. We find that model-free dark hole digging methods achieve comparable broadband contrasts to model-based methods, and highlight the calibration costs of model-free methods compared to model-based approaches. This study also reports the first time that electric field conjugation with the self-coherent camera has been applied for simultaneous multi-subband correction with a field stop. This study compares the advantages and disadvantages of each of these wavefront sensing and control algorithms with respect to their potential for future space telescopes.
\end{abstract}

\keywords{coronagraph, scalar vortex coronagraph, direct imaging, exoplanets, high contrast imaging}

{\noindent \footnotesize\textbf{*}Address all correspondence to Niyati Desai,  \linkable{ndesai2@caltech.edu} }



\section{INTRODUCTION}
\label{sec:intro}  
The primary mission objective outlined for NASA's proposed flagship mission, the Habitable Worlds Observatory (HWO), is to directly detect and acquire the spectra of 25 potentially habitable planets \cite{Astro2020_Report}. Coronagraphy is a promising solution for direct imaging in space, commonly employing a focal plane mask and pupil plane stop to suppress on-axis starlight and facilitate the imaging of off-axis planets. Despite major advancements over the past two decades, achieving the necessary contrast level of $10^{-10}$ in broadband light remains a challenge for current coronagraph technologies \cite{Ruane2022}.

A major obstacle hindering coronagraphs from achieving their theoretical starlight rejections is the presence of small aberrations that introduce starlight into the coronagraphic image, resulting in a speckle field. Deformable mirrors (DMs) offer a solution to address these speckles by correcting wavefront errors upstream of the coronagraphic mask. The role of DMs is to correct the wavefront to obtain a high quality point spread function (PSF) and to further suppress speckles, creating a dark hole region of high contrast in the final image plane suitable for exoplanet imaging.

Establishing a dark hole region at $10^{-10}$ contrast levels demands precise sensing and control of the electric field (E-field). Wavefront sensing methods that introduce non-common path optics increase the complexity of the instrument in this context. Instead, a control loop around the DMs and the final science image enables the estimation and control of the wavefront to create a dark hole through a process referred to as `dark hole digging' \cite{Malbet1995}. In recent years, various dark hole digging methods have emerged, holding potential applications for the HWO mission concept. 

This work focuses on three such methods. Here the terms `model-based' and `model-free' refer to whether a numerical simulated model of the optical system is necessary or not. The first is the strictly model-based dark hole digging technique known as Pairwise Probing with Electric Field Conjugation (PWP + EFC) \cite{Give'on2007}. The effectiveness of PWP + EFC relies on an accurate model of the coronagraph. Two model-free dark hole digging approaches are also considered in this paper: the self-coherent camera \cite{Baudoz2006} with EFC (SCC + EFC), and implicit electric field conjugation (iEFC) \cite{ruffio2022, Haffert_2023}. Wavefront sensing and control (WFSC) is broken down into a) sensing algorithms and b) control algorithms. Both sensing and control are required to correct wavefront aberrations. PWP and SCC are sensing algorithms while EFC is a control algorithm. iEFC is a combined sensing and control algorithm. Therefore, PWP or SCC can be paired with EFC, while iEFC is already a complete WFSC algorithm.

For HWO, a coronagraph achieving $10^{-10}$ contrasts necessitates not only highly precise and well-developed WFSC methods but also high-performance focal plane masks. The vortex coronagraph has been identified as a promising focal plane mask technology for the next space telescope \cite{LUVOIRReport, Juanola2022}. There are two flavors of vortex coronagraphs: the vector vortex coronagraph (VVC)\cite{Mawet2005} and the scalar vortex coronagraph (SVC), \cite{Ruane2019} which differ only in how they imprint phase into the beam. SVCs turn incoming wavefronts into an optical vortex through longitudinal phase delays, whereas VVCs use geometric phase shifts. While the VVC has so far demonstrated better contrast levels in laboratory settings, ongoing SVC design and testing indicate its potential to achieve comparable performance \cite{Galicher2020, Desai_2023, Desai_2024}. The predominant method for testing the performance of VVCs and SVCs in high-contrast imaging testbeds has been the model-based PWP + EFC. Doelman et al. (2023) did compare EFC and iEFC for a VVC experiment, however this comparison was implemented on different testbeds.\cite{Doelman2023} In this study, the model-free SCC + EFC and iEFC techniques are also implemented for testing both the SVC and VVC all on a single testbed. Notably, the these model-free approaches have not been applied to compare the performance of SVC and VVC before. 

This study aims to understand the differences in applying model-free versus model-based dark hole digging methods. In answering that question, this study has two main goals: 1) compare SVC and VVC performance under identical laboratory conditions and 2) identify the advantages and disadvantages of the three aforementioned WFSC algorithms when implemented in a single testbed. To guide the design of future space telescopes, a systematic comparison of these technologies is imperative. This study focuses on two key components: the focal plane mask and the WFSC algorithm.

For a standardized comparison, one testbed with fixed environmental conditions is necessary for conducting this study. The In-Air Coronagraph Testbed (IACT) at the Jet Propulsion Laboratory (JPL)\cite{Baxter2021} was employed for conventional model-based PWP + EFC and model-free iEFC, with modifications enabling the implementation of model-free SCC + EFC. This study not only focused on the average contrast in broadband light as a performance metric for these three methods but also considered convergence rates and how testbed stability influenced the results for both scalar and vector vortex coronagraphs. In Section~\ref{sec:svcvvc} the focal plane mask designs are introduced, Section~\ref{sec:techniques} covers the dark hole digging techniques, Section~\ref{sec:iact} explains the experimental setup, Section~\ref{sec:results} presents the narrowband and broadband results, and Section~\ref{sec:convergence} discusses the convergence rates observed of the techniques employed in this study.

\section{CORONAGRAPHIC FOCAL PLANE MASKS}
\label{sec:svcvvc}

The focal plane masks considered in this study are two kinds of vortex coronagraphs. The core principle behind a vortex coronagraph involves the rejection of on-axis starlight through the introduction of a spiral phase pattern onto the incoming wavefront. Placing a vortex mask at the focal plane induces an optical vortex phase ramp, $e^{il\theta}$, where $\theta$ represents the azimuthal angle, and $l$ denotes the topological charge, signifying the number of phase wraps. When a phase ramp with an even integer topological charge $l$ is multiplied by an on-axis airy function and propagated to the following pupil, the result is that all of the light is diffracted outside of the pupil\cite{Foo2005, Mawet2005}. Both flavors of vortex coronagraphs, namely the VVC and the SVC, apply the $e^{il\theta}$ phase ramp to the wavefront. However, the former utilizes a polarization-dependent geometrical phase shift, while the latter relies on a wavelength-dependent longitudinal phase delay.

One of the drawbacks of using a VVC is that an analyzer and polarizer are used to filter the light, allowing only a single polarization state to encounter the VVC mask resulting in a 50\% throughput loss. This is necessary because unpolarized light would result in the VVC imprinting two phase ramps of opposite helicity: $e^{+il\theta}$ and $e^{-il\theta}$. The two polarization states accumulate distinct aberrations which complicate wavefront control. Another limitation of the VVC is stellar leakage due to imperfect retardance. This leakage leads to a portion of the beam passing through the mask without acquiring the phase ramp which adds a faint, on-axis Airy pattern in the final image plane.

Meanwhile, less development and testing have been dedicated to SVCs. Current SVC focal plane masks exhibit chromaticity where the performance degrades at wavelengths deviating from the design central wavelength. Zeroth order leakage, or starlight which does not acquire an optical vortex phase after the focal plane mask, becomes a limiting factor in contrast performance \cite{Desai_2023}. Strategies to achromatize SVC designs involve varying the surface topography of the focal plane mask, combining materials with different refractive indices into a dual-layer focal plane mask, or employing metasurfaces optimized for broadband phase control \cite{Swartzlander2006, Ruane2019, Galicher2020, Desai_2022,Desai_2024, konig2023, Palatnick23}.

These two vortex coronagraph implementations present opposing advantages and disadvantages in the trade-off between polarization insensitivity and achromaticity, both crucial considerations for a future space telescope coronagraph instrument.  While the VVC and the SVC are two contending types of focal plane masks that offer effective starlight rejection, both require some technological advancements to achieve the desired $10^{-10}$ contrasts.



\section{DARK HOLE DIGGING TECHNIQUES}
\label{sec:techniques}
\begin{figure} [h]
    \begin{center}
    \includegraphics[width=0.9\textwidth]{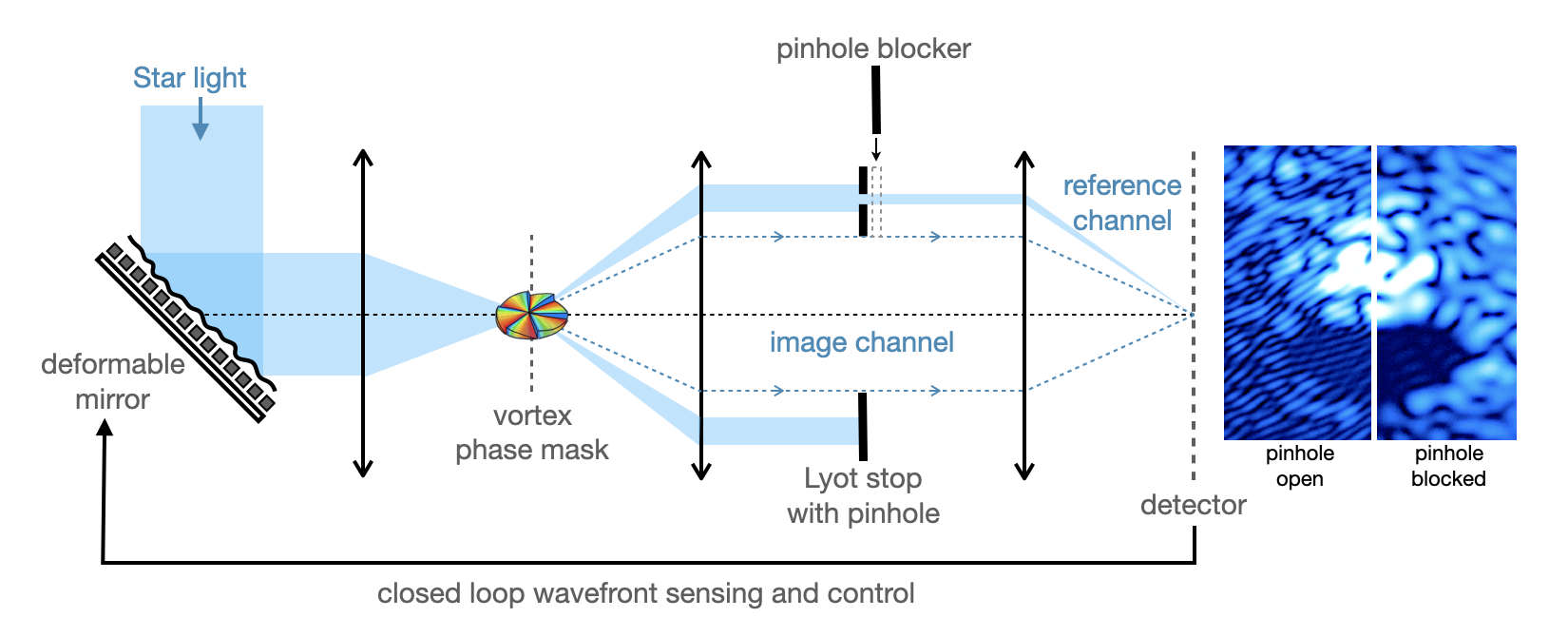}
    \end{center}
    \caption[fig:schema] 
    { \label{fig:schema} 
    Diagram of a coronagraph system equipped to perform wavefront sensing and control with conventional pairwise probing and electric field conjugation, self-coherent camera and electric field conjugation, as well as implicit electric field conjugation. Note the self-coherent camera technique uses the pinhole blocker to open and close the reference channel whereas the other two techniques do not use interference fringes in the final image and therefore do not require the pinhole in the Lyot stop to be open.
    }
\end{figure}

The dark hole digging algorithms examined in this paper can be categorized as either model-free or model-based, yet their common goal is to efficiently reduce the starlight within a specified region in the final focal plane. Figure~\ref{fig:schema} shows the optical system required for closed loop wavefront sensing and control with the three WFSC techniques presented in this paper. For all these methods, the science image is used for the wavefront sensing and the DM is used for the wavefront control. The right panel of Fig.~\ref{fig:schema} with the pinhole open shows an example final science image where we can see the speckle pattern of starlight through the coronagraph and the dark hole region where the speckles have been partially suppressed.

Despite the different naming conventions, all three methods discussed in this paper are methods of conjugating the electric field to reduce the residual starlight in the coronagraphic image. They are implemented with a DM upstream of the coronagraph in closed loop with the final focal plane camera acting as the sensor. The camera measures the intensity which is passed on the to the estimator which then tries to estimate the electric field of the residual starlight, or some proportional quantity. Finally the controller uses the estimate and determines the optimal commands to send to the DM to minimize the residual electric field. Typically a Jacobian matrix which describes the influence of the DM on the final focal plane image is required. This can either be inferred computationally, using a detailed optical model of the entire system, or experimentally, through systematically commanding a basis of shapes, or DM modes, on the DM and measuring the effect on the final image.

The general cost function followed by the controller is presented in Equation \ref{eq:costfunc}. It illustrates how EFC aims to effectively minimize the sum of a quantity proportional to the estimated E-field in the final science focal plane, $\rchi$, and the product of the DM commands, $u$, and the Jacobian, $G$. The variable $\rchi$ is intentionally retained as a placeholder, as this quantity varies among the three WFSC methods discussed in this paper. The objective of the algorithm is to determine the DM commands, $u$, that nullify the total E-field in the image plane.

\begin{equation}
\label{eq:costfunc}
J= min\left[ \lvert G u + \rchi \rvert ^{2} \right] 
\end{equation}

\begin{figure} [h]
    \begin{center}
    \includegraphics[height=5cm]{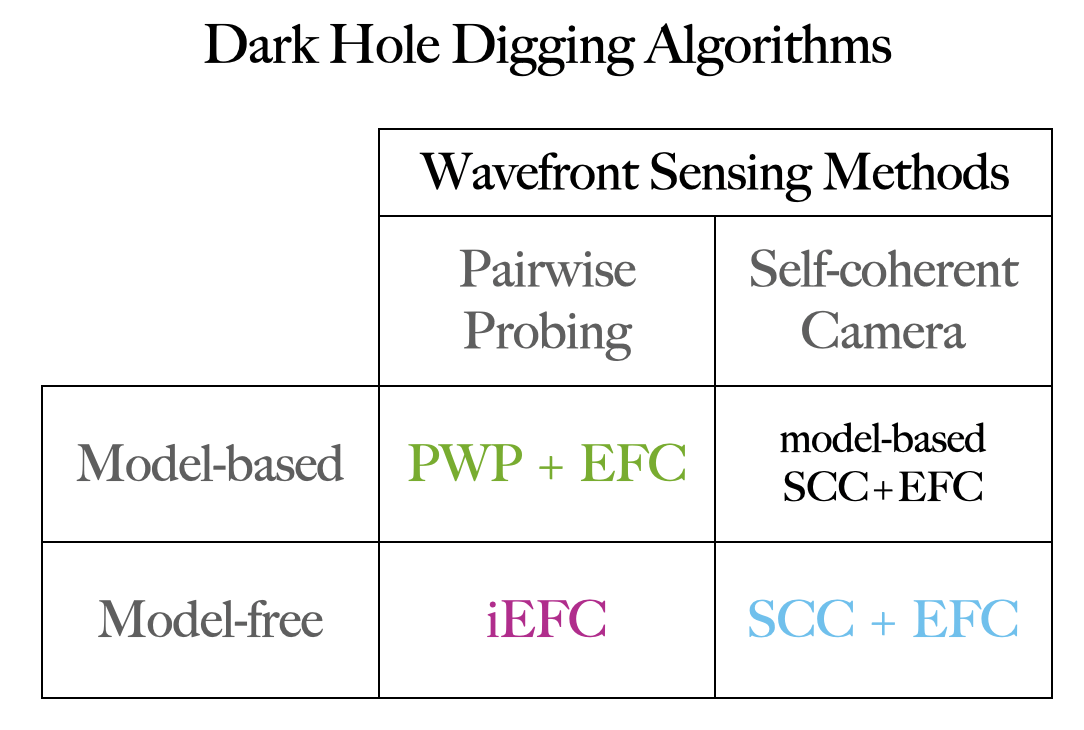}
    \end{center}
    \caption[fig:vvcprof] 
    { \label{fig:efc_matrix} 
    Matrix of various techniques of electric field conjugation split by their wavefront sensing methods and whether they require a system model. Pairwise probing (PWP) + electric field conjugation (EFC), self coherent camera (SCC) + electric field conjugation, and implicit electric field conjugation (iEFC) shown in the matrix are all dark hole digging methods implemented in this study. Note a model-based hybrid SCC has been partly investigated in Mazoyer et al. 2013\cite{Mazoyer2013}.
    }
\end{figure}

The matrix in Figure \ref{fig:efc_matrix} shows the relations between the three dark hole digging algorithms that were implemented in this study and an additional model-based SCC method which was not considered. They differ in implementation and wavefront sensing techniques, but more fundamentally in which quantity they are measuring and minimizing within the dark hole region.

The first technique discussed in this paper is conventional PWP + EFC which uses a model to estimate and directly minimize the E-field inside the dark hole $(E_{S})$. The second technique is SCC + EFC which introduces a pinhole in the pupil plane (seen in Figure~\ref{fig:schema}), then measures the intensity of the resulting interference fringes within the dark hole $(I_{-})$ and aims to minimize it. The third technique is iEFC, which uses pairwise probes on the DM to measure changes in intensity $(dI)$ in the final science image dark hole region and aims to minimize that change. Both SCC+EFC and iEFC operate without an optical model of the system. 

\subsection{Model-Based: Pairwise Probing + Electric Field Conjugation}
\label{subsec:pwpefc}

Conventional PWP + EFC is the algorithm used to obtain the best contrast demonstrated to date on the Decadal Survey Testbed at JPL \cite{Seo2019}. Introduced as simply EFC in 2007 by Give'on et al., EFC minimizes the sum of the estimated E-field in the final science plane and the E-field due to the corrective elements throughout the system\cite{Give'on2007}.

This model-based technique can be separated into a wavefront sensing technique (PWP) and a control algorithm (EFC). The wavefront sensing is done by applying pairwise sinc function probes (+/-) to the DM and measuring the final focal plane image. By applying known shapes on the DM which induce different electric fields at a location and recording the resulting intensity, the difference between these images can be calculated and used with the model of the entire optical system to determine the best estimate of the speckle electric field, $E_{S}$.

The control step involves first forming the Jacobian $G_{EFC}$ by moving each actuator of the DM in simulation and recording the induced focal plane E-field. The Jacobian is inverted and combined with the E-field estimate to iteratively compute the DM commands that minimize $E_{S}$ inside the correction region. Effectively Equation \ref{eq:costfunc} becomes: 
\begin{equation}
\label{eq:costfuncEFC}
J= min\left[ \lvert G_{EFC} u + E_{S} \rvert ^{2} \right] 
\end{equation}
where $G_{EFC} = \partial E_{S}/\partial u$ and can be created entirely from the model of the system. PWP + EFC would not require any images per mode to build the Jacobian but a minimum of two probes hence four images per iteration. For broadband light, the bandwidth is divided into multiple subbands spanning the entire bandpass and the required exposures for this technique would apply for each subband. 

\subsection{Model-Free: Self-Coherent Camera + Electric Field Conjugation}
\label{subsec:scc}

The concept of the classical SCC was conceived by Baudoz et al.~in 2006\cite{Baudoz2006}. Since then it has been mainly implemented on the Très Haut Dynamique bench (THD) at the Paris Observatory\cite{Baudoz2018,Potier2020} which demonstrated contrast performance around $10^{-9}$ in-air with a Four Quadrant Phase Mask coronagraph\cite{Rouan2000}.

Like the conventional PWP + EFC, this model-free dark hole digging technique can also be broken into a wavefront sensing method (SCC) and a control algorithm (EFC). Wavefront sensing with the SCC works by introducing a pinhole in the Lyot plane at a location where coherent starlight rejected by the coronagraphic mask lands\cite{Galicher2010}. This pinhole creates a reference channel which interferes with any residual starlight leaking through the image channel, thereby creating fringes at the final focal plane (see Figure~\ref{fig:schema}). In principle, this fringed image contains all the information needed to estimate the unwanted E-field in the coronagraphic image. The fringe intensity ($I_{-}$), which is proportional to the speckle E-field, is then minimized in the control step.

The control step follows almost the same EFC procedure when paired with SCC as with PWP, except the Jacobian ($G_{SCC}$) is data-driven. Effectively Equation \ref{eq:costfunc} becomes: 
\begin{equation}
\label{eq:costfuncSCC}
J= min\left[ \lvert G_{SCC} u + I_{-} \rvert ^{2} \right] 
\end{equation}
where $G_{SCC} = \partial I_{-}/\partial u$ and is created by actuating the DM and measuring the SCC response as explained above. After systematically building this matrix, the estimate from the wavefront sensing and the Jacobian are combined to iteratively control the DM to minimize $I_{-}$ inside the dark hole. SCC + EFC would require a minimum of two images (+/-) per mode to build the Jacobian and one image per iteration for each subband.

There are several variations of SCC wavefront sensing which can be applied to obtain the $I_{-}$ mentioned above. The classical SCC relies on the Fourier transform of the fringed image to get an estimate for the final E-field\cite{Galicher2010,Mazoyer2013,Delorme2016}. Building off the classical SCC, two simple modifications further improve its wavefront sensing capabilities and were implemented in this study for broadband experiments. The first variation is based on Thompson et al.~(2022)\cite{Thompson2022}. Instead of using the Fourier transform to extract the fringe intensity, the closed pinhole image (unfringed) is subtracted from the opened pinhole image (fringed) to retain the fringe intensity $I_{-}$. Although the wavefront sensing is improved, this variant takes longer to build the Jacobian since two images are required to estimate the E-field where classical SCC needs only one. This variant allows for the use of a field stop, which poses a significant challenge in classical SCC. This is because classical SCC relies on the Fourier Transform which would be distorted by introducing a field stop. This variant also offers the ability to use pinholes closer to the science channel which potentially extracts more light for wavefront sensing\cite{Martinez2019}. This capability will be tested in a future work. 



The second variation on SCC opens multiple pinholes instead of just one. This multi-reference SCC method is based on Delorme et al. (2016)\cite{Delorme2016}. With a single pinhole, the fringe contrast is reduced due to chromatic blurring away from the central fringe. With three pinholes in the Lyot stop, all at the same radius away from the center, overlapping fringes are created in different directions in the coronagraphic image, increasing the area of high contrast fringes. Delorme et al. demonstrated that such a design can be used to enhance SCC broadband capabilities. Hence, for this study it was implemented to effectively increase the amount of light in the reference channel and allow for improved wavefront sensing across the 10\% bandwidth. Desai et al.~(2023) provides an experimental comparison between these SCC variants and shows that the three pinhole technique without the Fourier transform demonstrates the most efficient WFSC of the broadband SCC methods\cite{Desai2023SPIE_SCC}.

\subsection{Model-Free: Implicit Electric Field Conjugation}
\label{subsec:iefc}
Implicit electric field conjugation was presented formally by Haffert et al.~in 2023\cite{Haffert_2023} and has recently become more commonly implemented on high contrast imaging benches for its appeal as a model-free version of conventional PWP+EFC. Unlike the other two dark hole digging methods presented in this paper, iEFC cannot be split into wavefront sensing and control, as the two occur concurrently. 

The first step of iEFC involves applying a set of (+/-) modes to the DM with superimposed (+/-) pairwise probes. In this case, Fourier modes were used. The difference in intensity, ($dI$), from the probe measurements is used to build the Jacobian, $G_{iEFC}$. Then, identical pairwise probes are applied to the DM at each iteration to measure the difference in intensity ($dI$) in the correction region. This quantity, proportional to the speckle E-field, is therefore minimized when multiplied with the inverse of the initially calibrated Jacobian. Effectively Equation \ref{eq:costfunc} becomes 
\begin{equation}
\label{eq:costfunciEFC}
J= min\left[ \lvert G_{iEFC} u + dI \rvert ^{2} \right] 
\end{equation}

where $G_{iEFC} = \partial(dI)/\partial\alpha$. One of the unique advantages of iEFC is that it requires no additional optics or hardware since the primary wavefront sensing technique is PWP involving the DMs. iEFC would require a minimum of two probes, so eight images per mode to build the Jacobian and four images per iteration for each subband. 

\section{EXPERIMENTAL SETUP}
\label{sec:iact}

\begin{figure} [t]
    \begin{center}
    \includegraphics[height=7.2cm]{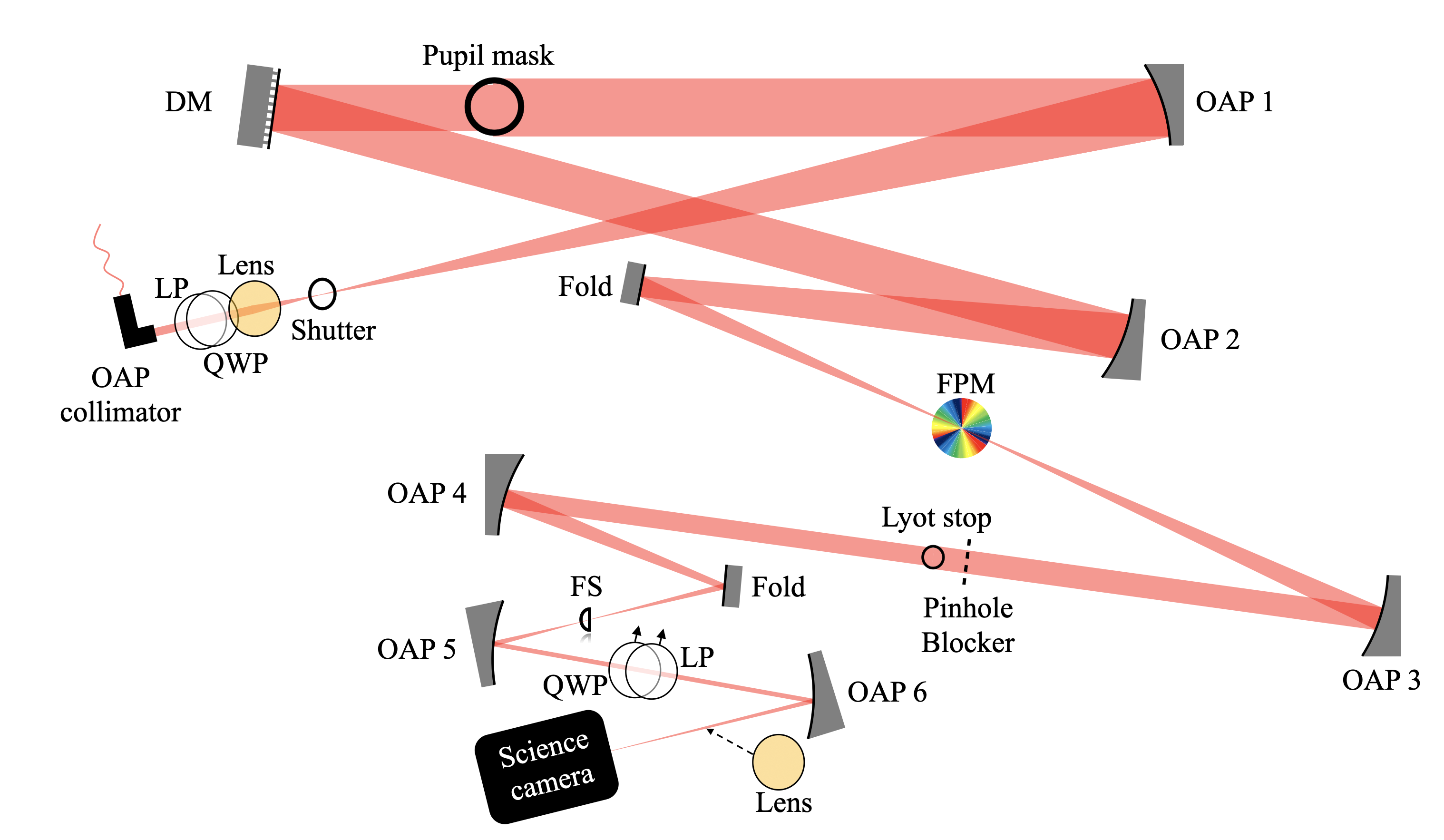}
    \end{center}
    \caption[fig:iact] 
    { \label{fig:iact} 
    Schematic of the In-Air Coronagraph Testbed at JPL.
    }
    \end{figure}

In this study, dark holes were dug on the In Air Coronagraph Testbed (IACT) at JPL, shown in Figure~\ref{fig:iact},\cite{Baxter2021} for both the VVC and SVC in uniform conditions in broadband light using all three WFSC algorithms. To maintain as many of the experimental conditions constant as possible, the same bandwidths and subbands were used: $10\%$ bandwidth split into three subbands. However the coronagraphic focal plane masks on IACT have different designed central wavelengths. The VVC is a liquid crystal charge 6 polarization dependent vector vortex mask designed with $\lambda_0 = 635$~nm and and the SVC is a charge 6 wrapped staircase scalar vortex wavelength dependence mask designed with $\lambda_0 = 775$~nm.  Note that the intensity of the source, an NKT SuperK supercontinium laser with Varia tunable filter, decreases significantly for wavelengths greater than 650~nm, and has an upper limit of 800~nm. As a result, the 10\% broadband VVC measurements were centered at $\lambda_{0}=635$ nm and the SVC measurements were centered at $\lambda_{0}=760$~nm. Consequently the overall flux for the SVC experiments is fainter than for the VVC experiments, making wavefront sensing with SCC more challenging. This is not related to the FPM mask or the WFSC algorithm, but instead  due to the continuum light source used on IACT. To account for this, the broadband contrast is defined here as the averaged intensity of the coronagraphic subband images divided by the peak intensity of the unocculted broadband pseudostar image.

IACT has already been used to conduct PWP + EFC for both the VVC and SVC\cite{Desai_2023}. The VVC and SVC are both mounted on a stage with a Newport LTA actuator, so consecutive experiments can quickly switch between masks while maintaining similar laboratory conditions\cite{Baxter2021}.  In order to implement the SCC, only two modifications in the pupil plane were required: a new Lyot stop with multiple pinholes to create the reference channels, and a pinhole blocker in front of the Lyot mask to open and close the pinholes while keeping the Lyot stop open. 

Note that the SCC Lyot stop was designed to have 3 outer pinholes (used for this study) and 3 inner pinholes (for a possible alternative SCC experiment). The pinhole size was optimized for the VVC central wavelength and their positions were chosen to be within in the VVC `ring of fire' (a ring of rejected starlight created by the FPM in the Lyot plane) so that the maximum amount of coherent starlight is transmitted through this reference channel. The Lyot stop design used in this study is therefore slightly sub-optimal for the SVC coronagraph. In addition to the custom Lyot stop, the custom pinhole blocker was designed to keep the Lyot stop central channel open while blocking or unblocking various sets of pinholes in a controlled manner. Further details on these upgrades can be found in Desai et al.~(2023)\cite{Desai2023SPIE_SCC}. Additionally, a preliminary comparison in narrowband light of different SCC variations with this pinhole blocker using one versus three pinholes simultaneously or using the Fourier transform method is reported in Desai et al. (2023)\cite{Desai2023SPIE_SCC}.

Both the VVC and SVC on IACT are charge 6 vortex coronagraphs with an inner working angle of 3~$\lambda/D$. For this reason, the dark hole digging algorithms tested in this study targeted a D-shaped correction region from 3 - 10 $\lambda_0/D$. Trials across all three WFSC algorithms were run for 50 iterations. In addition, an oversized D-shaped field stop was used just before the science camera to block the center of the PSF within the inner working angle and prevent any ghost reflections from reaching the camera.

\section{RESULTS}
\label{sec:results}

\begin{figure} [t]
    \begin{center}
    \includegraphics[width=\textwidth]{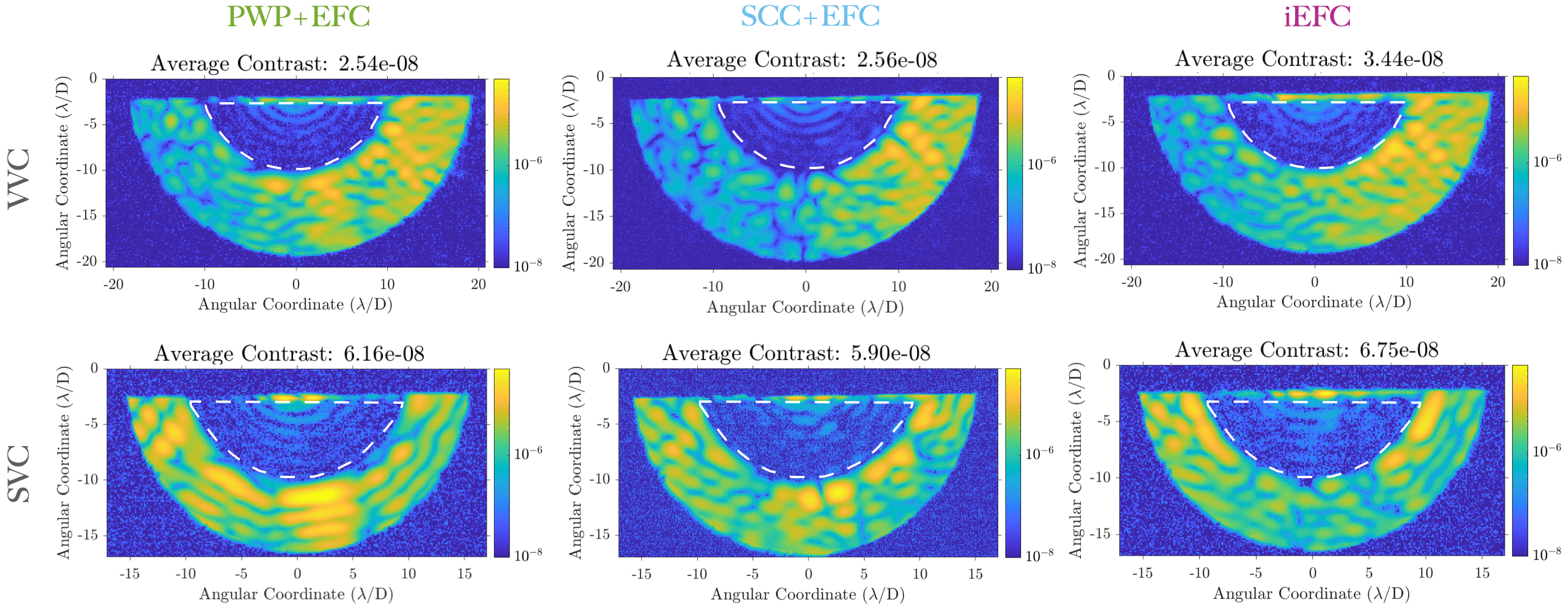}
    \end{center}
    \caption[fig:dhsnarrowband] 
    { \label{fig:narrowband_dhs} 
    Narrowband 2\% dark holes and the average contrast within the scoring region from 3-10 $\lambda/D$ with the vector vortex coronagraph (top row) and scalar vortex coronagraph (bottom row) from the In-Air Coronagraph Testbed with three WFSC algorithms. The VVC measurements were taken at 635 nm and the SVC measurements were at 775 nm.
    }
\end{figure}

Figure~\ref{fig:narrowband_dhs} shows the 2\% narrowband dark holes dug by all three techniques and their average contrast within a scoring region from 3-10 $\lambda/D$ (in a dashed white line). Note that the VVC trials were taken at a shorter wavelength than the SVC trials, but the size of the dark hole remains the same in units of $\lambda/D$. In narrowband light with both the VVC and the SVC, the six dark holes presented here all show similar average contrasts roughly between 2-7 \texttimes \num{e-8} within the specified scoring region.

\begin{figure} [t]
    \begin{center}
    \includegraphics[width=\textwidth]{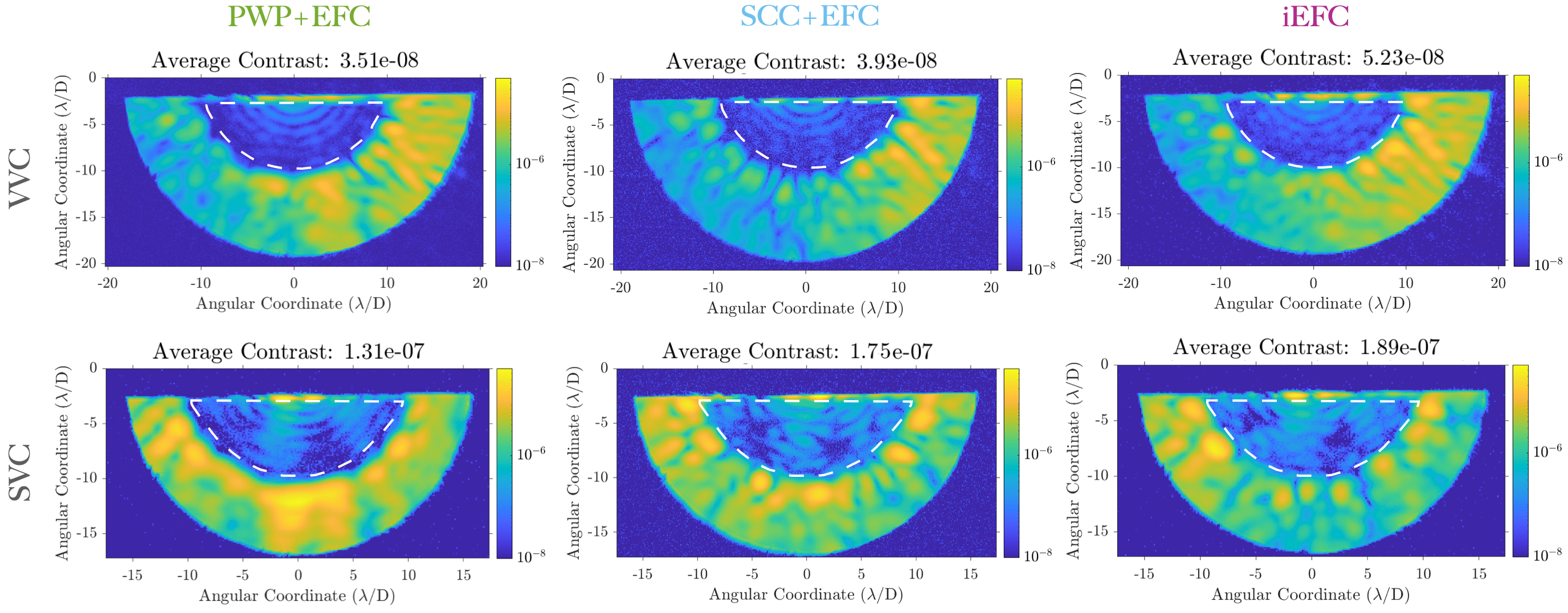}
    \end{center}
    \caption[fig:dhsbroadband] 
    { \label{fig:broadband_dhs} 
    Broadband 10\% dark holes and the average contrast within the scoring region from 3-10 $\lambda/D$ with the vector vortex coronagraph (top row) and scalar vortex coronagraph (bottom row) from the In-Air Coronagraph Testbed with three WFSC algorithms. The VVC measurements were centered at 635 nm and the SVC measurements were centered at 760 nm.
    }
\end{figure}

Meanwhile a clear distinction between the average contrasts for the VVC and SVC is visible in broadband light. Figure~\ref{fig:broadband_dhs} shows the 10\% broadband dark holes dug by all three techniques and their average contrast within the same scoring region from 3-10 $\lambda/D$ (in a dashed white line). In broadband, the VVC average contrasts were all roughly \num{4e-8} whereas the SVC average contrasts were all generally a factor of 3 worse than the VVC averages, roughly \num{1e-7}. The consistent performance of the SVC across all three wavefront sensing and control techniques, reaching approximately an order of magnitude worse contrasts than the VVC in 10\% broadband light, indicates a model-mismatch is not the limitation of the SVC. 

Additionally between the vector and scalar vortex coronagraphs, aside from the hardware defects mentioned in Section~\ref{sec:iact}, a large difference in average contrasts is not present between the narrowband dark holes. However a difference is apparent between the broadband results, and is evidence that the SVC contrasts are limited by chromaticity. The VVC dark holes exhibit a clearer ringing pattern compared to the SVC dark holes which show a much less prominent trend. Both the VVC polarization leakage and SVC chromatic leakage would be in the shape of an Airy pattern in the final dark hole and yield profiles like those observed.

{\renewcommand{\arraystretch}{1.6}
\begin{table} [t!]
\begin{center}
\resizebox{\textwidth}{!}{
    \begin{tabular}{rcccc}
    \toprule
        & \multicolumn{2}{c}{Vector Vortex} & \multicolumn{2}{c}{Scalar Vortex} \\
         & 2\% & 10\% & 2\% & 10\% \\
        \cmidrule(r){2-3} \cmidrule(r){4-5}
         PWP + EFC & $2.54 (\pm 1.6) \times 10^{-8}$ & $3.51 (\pm 3.3) \times 10^{-8}$ & $6.16 (\pm 7.4) \times 10^{-8}$ & $1.31 (\pm 1.7) \times 10^{-7}$\\ 
         SCC + EFC & $2.56 (\pm 1.9) \times 10^{-8}$ & $3.93 (\pm 2.7) \times 10^{-8}$ & $5.90 (\pm 5.9) \times 10^{-8}$ & $1.75 (\pm 1.5) \times 10^{-7}$\\ 
         iEFC & $3.44 (\pm 2.4) \times 10^{-8}$ & $5.23 (\pm 3.0) \times 10^{-8}$ & $6.75 (\pm 5.2) \times 10^{-8}$ & $1.89 (\pm 1.9) \times 10^{-7}$\\ 

         \bottomrule
    \end{tabular}
    }
        \vspace{5 mm} 
    \caption{\label{tab:contrasts} Summary table of best dark hole normalized intensity with standard deviations from 3-10 $\lambda/D$ for the scalar vortex coronagraph and vector vortex coronagraph in 2\% narrowband and 10\% broadband light with pairwise probing and electric field, self-coherent camera with electric field conjugation, and implicit electric field conjugation.}
\end{center}
\end{table}
}

Table~\ref{tab:contrasts} shows a summary of all the average contrasts achieved for both the VVC and SVC obtained with PWP + EFC, SCC + EFC, and iEFC. Overall both the 2\% narrowband and the 10\% broadband dark holes were consistent across WFSC techniques. Since the dark holes across all three techniques are in agreement, these results indicate the average contrast levels are not limited by the WFSC method. Instead the VVC is likely limited by polarization leakage due to the analyzer/polarizer and the SVC is likely limited by the focal plane mask's chromatic zeroth order leakage,which is in agreement with the simulated SVC performance\cite{Desai_2023}.

\section{CONVERGENCE RATES}
\label{sec:convergence}

One of the primary objectives of this study was not only to use three different dark hole digging methods in one environment and compare the best contrasts achieved, but also to consider the implementations of each and note their advantages and disadvantages.

\begin{figure} [t!]
    \begin{center}
    \includegraphics[width=0.8\textwidth]{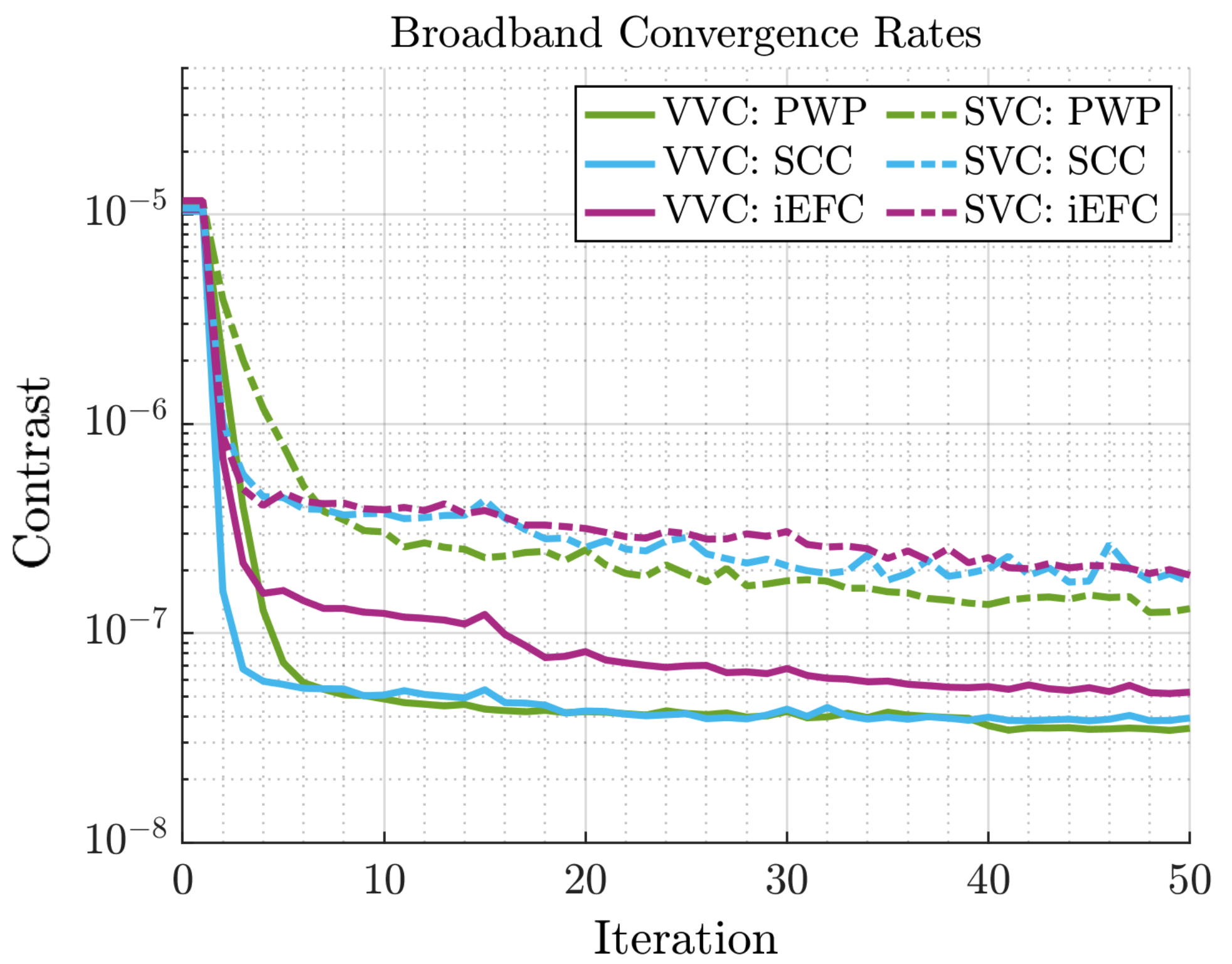}
    \end{center}
    \caption[fig:convergence] 
    { \label{fig:convergence} 
    Convergence rates of three WFSC algorithms: pairwise probing (PWP) and electric field conjugation (EFC) (green), self-coherent camera (SCC) with electric field conjugation (blue), and implicit electric field conjugation (iEFC) (pink), for both the vector vortex coronagraph (VVC) (solid lines) and scalar vortex coronagraph (SVC) (dotted lines) in a 10$\%$ bandwidth plotted against iteration.
    }
    \end{figure}

Figure~\ref{fig:convergence} shows the convergence rate of each of the three techniques in 10\% broadband light. The trials shown here yielded the best average broadband contrasts within the 3-10 $\lambda/D$ dark hole. The dashed lines are SVC trials and the solid lines are VVC trials. The PWP + EFC trial is shown in green, the iEFC in pink, and the SCC in blue.

The model-free methods reach deeper contrasts in the first few iterations than the model-based method. The faster convergence rate for both SCC + EFC and iEFC are clearly visible for both the SVC and VVC compared to the PWP + EFC. It was expected that the SCC sensitivity would outperform PWP + EFC and iEFC. This is because the SCC is only limited by the number of photons going through the pinholes, which can be increased with more cleverly designed coronagraphs or Lyot stops\cite{Delorme2016,Martinez2019,Gerard2020,Haffert22}. On the other hand, the PWP + EFC and iEFC algorithms assume a linear approximation that restricts the probe amplitude and therefore the amount of leaked photons used for an accurate E-field estimation\cite{Groff2016}.

\begin{figure} [t!]
    \begin{center}
    \includegraphics[width=\textwidth]{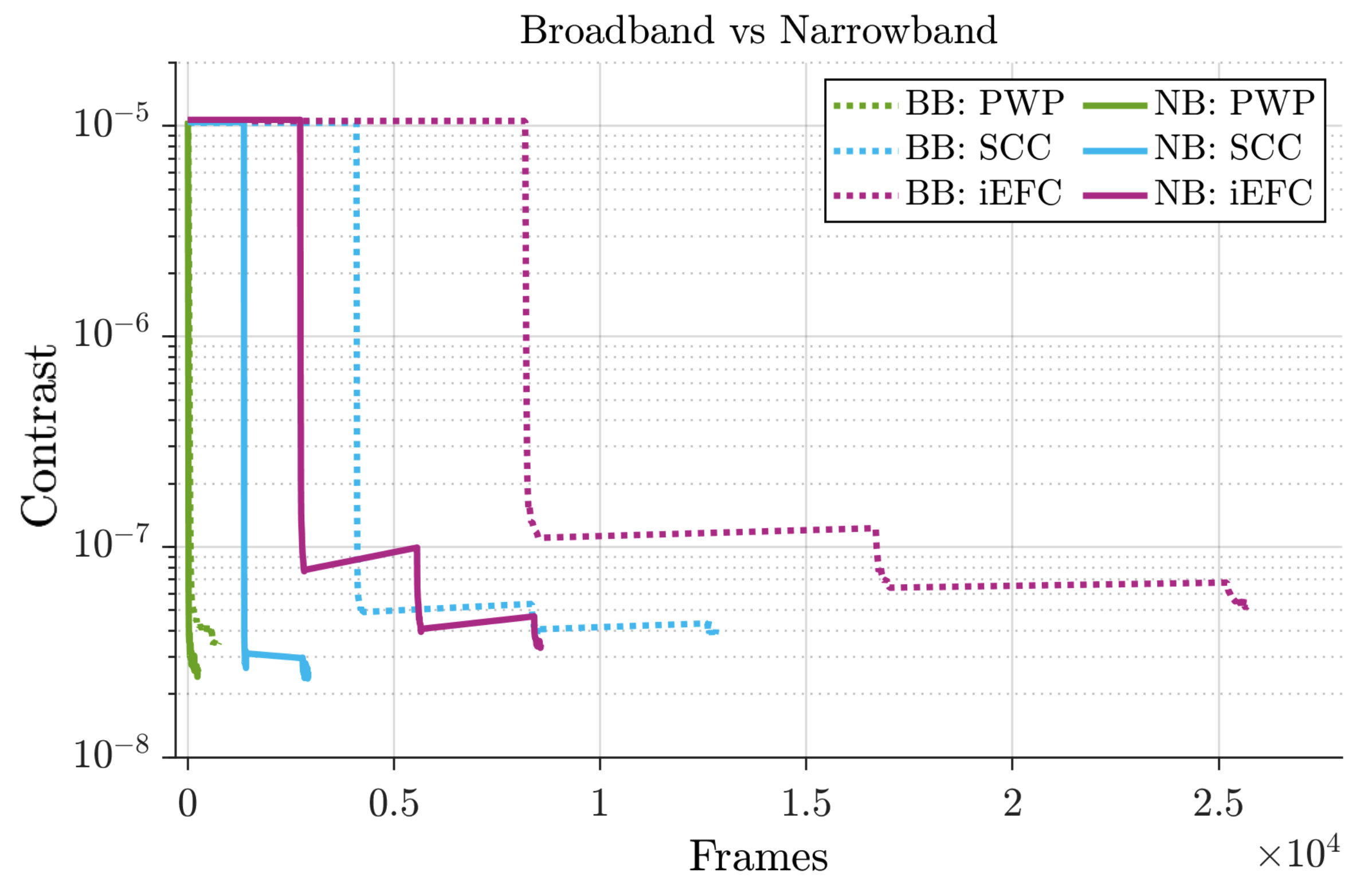}
    \end{center}
    \caption[fig:convergence] 
    { \label{fig:frames_convergence} 
    Convergence rates of three WFSC algorithms: pairwise probing (PWP) and electric field conjugation (EFC) (green), self-coherent camera (SCC) with electric field conjugation (blue), and implicit electric field conjugation (iEFC) (pink), for both 2\% narrowband (NB) light (solid lines) and 10\% broadband (BB) light} (dotted lines) plotted against number of frames, or total exposures.
    
    \end{figure}

When the same trials are plotted against number of exposures, or frames, and include the cost required on IACT to build the Jacobian without any model, the model-free methods no longer seem to be more efficient. Figure~\ref{fig:frames_convergence} shows the long period at the start of a trial used to build the Jacobian as well as, for both SCC and iEFC, the periods of relinearization during the trial. These empirical processes take on the order of a few hours, strongly dependent on the targeted contrast and stellar flux, and they are required for the model-free techniques before the first iteration as well as during the correction procedure. In this case, as seen in Figure~\ref{fig:frames_convergence}, the relinearization process is done twice more over the 50 iteration period. We do not consider here the time for DM registration, or model calibration for PWP + EFC which can be parallelized and depend on computational power. Additionally we don't include the characterization of the DM with a Zygo interferometer, a time-consuming process that would be required regardless of which algorithm is used. 

In general, the long runtimes for the model-free WFSC techniques would be sensitive to testbed instability and PSF drift during the long calibration periods. Throughout these experiments, we faced instabilities in the high bay leading to variation of final results, mainly for the model free algorithms. It's possible that this might be avoided on a vacuum testbed, improving measurements by maintaining stable conditions during the calibration procedures required in these experiments. However, the number of frames required for each of the steps in the model-free techniques would not change and this might be a limiting factor. 

\section{CONCLUSION}
\label{sec:conc}

In conclusion, this study found iEFC and SCC + EFC are both high performing model-free dark hole digging methods and ultimately reach equally high broadband contrasts (around \num{3e-8} for the VVC and around \num{1e-7} for the SVC in 10\% bandwidth) as model-based PWP + EFC. Through the course of implementing these three techniques on IACT, this study reports the first time SCC + EFC has been applied for simultaneous multi-subband correction with a field stop. 

A lot can still be optimized with the hardware installed for the SCC on IACT. In particular, the SCC Lyot stop pinhole geometry is optimized for the VVC instead of the SVC. The pinhole placement matches the theoretical `ring of fire' where the VVC mask throws light outside of the Lyot stop radius in the pupil plane, and could be redesigned to better fit the SVC's theoretical `ring of fire' for potentially improved performance. Although this study compares the performance of PWP + EFC, SCC, and iEFC directly, it's important to note these three dark hole digging algorithms have not all been fully studied and developed equally. PWP + EFC is the most established of these methods and the code architectures used to implement SCC + EFC and iEFC were both derived from one designed specifically for application to PWP + EFC. 

The results show that compared to VVCs, SVC performance is still limited by its chromaticity and likely not by model-mismatch in PWP + EFC. This work definitely motivates the need for better scalar vortex designs. Some ongoing efforts to achromatize scalar vortex coronagraphs include employing hybrid combinations of radial and azimuthal features or utilizing metasurface implementations \cite{Desai2023SPIE_Roddier, Desai_2024, konig2023, Palatnick23}. Both are strategies which yield complex focal plane masks that might benefit from model-free WFSC techniques. In general, this study found that model-free techniques are as effective in creating a dark hole with high contrast levels as model-based techniques, however indeed currently, model-free techniques take more time to converge due to the measurement of the Jacobian. They might be a preferred choice to deal with more complex mask designs like phase-induced amplitude apodization (PIAA) masks, metasurface scalar vortex phase masks or other future increasingly complex mask designs where model mismatch impedes performance\cite{Fogarty2022,konig2023,Palatnick23}.

Another conclusion of this study is that model-free methods require significantly longer time for Jacobian building. These techniques would be limited by testbed instabilities\cite{Leboulleux2016}, however might still be appropriate when no resources can be allocated to make a detailed optical model. In terms of iterations, the model-free methods sometimes converge faster and in the case of SCC + EFC, require less images per iteration. But in terms of number of exposures, the conventional model-based method saves a lot of calibration cost by using a model of the optical system instead of having to tediously poke each Fourier mode on the DM to build the Jacobian. Model-free techniques could be improved with a clever choice of mode basis to reduce the overall number of frames required to build the Jacobian. Or perhaps a combination of these methods using the SCC for wavefront sensing and an optical model to compute the DM correction, might offer a balance of the advantages offered by model-free and model-based WFSC techniques.


\subsection*{Disclosure}
This paper is an extension of the work submitted in the SPIE Proceedings: Desai et al. 2023.

\subsection*{Code and Data Availability}
All data in support of the findings of this paper are available within the article or as supplementary material.

\acknowledgments 
 This work was supported by the NASA ROSES APRA program, grant NM0018F610. Part of this research was carried out at the Jet Propulsion Laboratory, California Institute of Technology, under a contract with the National Aeronautics and Space Administration (80NM00018D0004).

\appendix    


\bibliography{report}   
\bibliographystyle{spiejour}   





\end{document}